\documentclass[conference]{IEEEtran}
\IEEEoverridecommandlockouts
\usepackage{booktabs}
\usepackage{multirow}
\usepackage{textcomp}
\usepackage{xcolor}
\usepackage{cite}
\usepackage{latexsym}
\usepackage{graphicx}
\usepackage{subcaption}
\usepackage{amsfonts,amssymb,amsmath}
\usepackage{nccmath}
\usepackage{optidef}
\usepackage{hyperref}
\usepackage{color,soul}
\usepackage{array}
\usepackage{algorithm}
\usepackage{algorithmic}
\usepackage[T1]{fontenc}
\usepackage[utf8]{inputenc}
\usepackage{fancyhdr}
\usepackage{lastpage}
\usepackage{caption}
\usepackage{soul}
\usepackage{subcaption}
\usepackage{setspace}
\usepackage{bm}
\def\BibTeX{{\rm B\kern-.05em{\sc i\kern-.025em b}\kern-.08em
    T\kern-.1667em\lower.7ex\hbox{E}\kern-.125emX}}
\usepackage{fancyhdr}
\fancypagestyle{firststyle}{
	\fancyhf{}
	\fancyhead[L]{H. Poddar, T. Yoshimura and A. Ishii "Validation of 3GPP TR 38.901 Indoor Hotspot Path Loss Model Based on Measurements Conducted at 6.75, 16.95, 28, and 73 GHz for 6G and Beyond'', \textit{Accepted in 2025 IEEE 101st Vehicular Technology Conference (VTC2025-Spring)}, Oslo, Norway, June 2025, pp. 1--6.}

}

\begin{document}
\title{
Validation of 3GPP TR 38.901 Indoor Hotspot Path Loss Model Based on Measurements Conducted at 6.75, 16.95, 28, and 73 GHz for 6G and Beyond

}
\author{\IEEEauthorblockN{Hitesh Poddar, Tomoki Yoshimura and Art Ishii}
\IEEEauthorblockA{Sharp Laboratories of America, Vancouver WA, USA, \{poddarh, yoshimurat, ishiia\}@sharplabs.com\\}
}
\maketitle

\thispagestyle{firststyle}

\begin{abstract}
This paper presents a thorough validation of the Third Generation Partnership Project (3GPP) Technical Report (TR) 38.901 indoor hotspot (InH) path loss model, as part of the 3GPP Release 19 study on “Channel model validation of TR 38.901 for 7-24 GHz,” for 6G standardization. Specifically, we validate the 3GPP TR 38.901 path loss model for the InH scenario in both line of sight (LOS) and non line of sight (NLOS) channel conditions, using the floating intercept (FI) and alpha-beta-gamma (ABG) path loss models. The validation focuses on specific frequencies, including 6.75 GHz and 16.95 GHz, as well as the broader 7-24 GHz and 0.5-100 GHz frequency ranges. The validation is based on real-world measurements conducted at 6.75 GHz, 16.95 GHz, 28 GHz, and 73 GHz by NYU WIRELESS using a 1 GHz wideband time domain based sliding correlation channel sounder in the InH scenario for both LOS and NLOS channel conditions. Our results confirm that the 3GPP TR 38.901 path loss model for the InH scenario remains valid for the 7–24 GHz range in both LOS and NLOS conditions and provide valuable input for 6G standardization efforts.

\end{abstract}

\section{Introduction}
The rapid evolution of 6G wireless communications has intensified global focus on the 7-24 GHz frequency range, commonly known as the FR3 band or the upper-mid band \cite{kang:2023:cellular-wireless-networks, jiang:2021:the-road-towards}. Frequently referred to as the "golden band," this frequency range offers an ideal balance between coverage and capacity \cite{bjornson:2024:enabling-6g-performance, semaan:2023:6g-spectrum-enabling}. At the International Telecommunication Union (ITU) World Radiocommunication Conference (WRC) held in 2023, several key frequency bands within the 7-24 GHz frequency range were identified for further study and harmonization under the International Mobile Telecommunications (IMT) 2030 framework or 6G \cite{ghosh:2024:world-radiocommunications}. Specifically, the ITU WRC in 2027 will examine the potential use of the following bands: 4.4-4.8 GHz, 7.125-7.250 GHz, 7.75-8.40 GHz, and 14.8-15.35 GHz \cite{ghosh:2024:world-radiocommunications}. Notably, the 7.125-8.4 GHz and 14.8-15.35 GHz bands have emerged as the potential bands of global harmonization for IMT 2030 \cite{ghosh:2024:world-radiocommunications, ghosh:2024:the-national-spectrum, semaan:2023:6g-spectrum-enabling, davidson:2024:national-spectrum}. 
Furthermore, at the ITU WRC in 2027 \cite{gsma:2024:the-road}, key decisions on global 6G spectrum allocations will be made, and it is expected that industry, academic institutions \cite{shakya:2024:comprehensive-fr1, shakya:2024:propagation, shakya:2024:wideband-penetration, chen:2024:an-experimental}, and other stakeholders \cite{katwe:2024:cmwave} will intensively study the 7-24 GHz frequency range \cite{zhang:2024:new-mid-band}. 
\par  Major global standardization bodies 
are actively studying the 7-24 GHz frequency range \cite{nga:2024:channel-measurements-and-modeling}. To support future cellular deployments in this frequency range, 3GPP approved a Release 19 study on ``Channel model validation of TR 38.901 for 7-24 GHz'' \cite{3gpp:2023:sid-channel-modelling}. Although 3GPP’s TR 38.901 channel model \cite{3GPP:2020:study-on-channel-model} currently spans the frequency range of 0.5-100 GHz, the data used to construct this channel model predominantly focus on FR1 frequencies (below 6 GHz) and FR2 frequencies (above 24 GHz), with limited data available for the 7-24 GHz frequency range \cite{3gpp:2023:sid-channel-modelling}. As a result, interpolation was primarily used to estimate the radio propagation characteristics within the 7-24 GHz frequency range based on measurement data at FR1 and FR2 frequencies \cite{3gpp:2023:sid-channel-modelling}. Thus, the 7-24 GHz frequency range, which lies between FR1 and FR2 frequencies, requires further study and validation to ensure that 3GPP TR 38.901 accurately captures the radio propagation characteristic for 7-24 GHz frequency range \cite{3gpp:2024:views-on-channel-model-sharp-nyu,3gpp:2024:views-on-channel-model-sharp}. \par Hence, in this paper, we validate the 3GPP TR 38.901 path loss model for the InH scenario in both LOS and NLOS channel conditions by deriving the FI and ABG path loss model parameters \cite{3gpp:2023:sid-channel-modelling} for specific frequencies such as 6.75 GHz and 16.95 GHz\footnote{6.75 GHz and 16.95 GHz was selected due to equipment constraints and Federal Communications Commission (FCC) authorization for conducting indoor and outdoor measurements in New York City, USA \cite{shakya:2024:comprehensive-fr1}.}, as well as for the entire 7-24 GHz and 0.5-100 GHz frequency ranges. 
The validation leverages real-world measurements at 6.75 GHz \cite{shakya:2024:comprehensive-fr1, rappaport:2024:point-data-for-site-specific}, 16.95 GHz \cite{shakya:2024:comprehensive-fr1, rappaport:2024:point-data-for-site-specific}, 28 GHz \cite{maccartney:2015:indoor-office-wideband}, and 73 GHz \cite{maccartney:2015:indoor-office-wideband}, conducted by NYU WIRELESS. While additional frequency points across the 7–24 GHz and 0.5–100 GHz ranges would improve the robustness of the validation; this was not feasible due to hardware limitations and the specific InH environments available during the measurement campaign. Nonetheless, the selected frequencies 6.75 GHz and 16.95 GHz are representative of the lower and upper ends of the 7–24 GHz band, offering meaningful insight into path loss behavior across this wideband. It is important to emphasize that the findings presented in this work should be considered as one of many contributions submitted to the ongoing 3GPP Release 19 efforts aimed at validating the 3GPP InH path loss model. The
key contributions of this paper are:
\begin{itemize}
    \item This paper contributes to ongoing 6G standardization efforts in 3GPP Release 19 by offering critical insights on InH path loss characteristics.
    \item The FI path loss model parameters are derived for 6.75 GHz and 16.95 GHz in both LOS and NLOS channel conditions for the InH scenario and comparisons are made with the 3GPP path loss model.
    \item The ABG path loss model parameters are derived for the 7–24 GHz and 0.5-100 GHz frequency ranges using measurements conducted at 6.75 GHz, 16.95 GHz, 28 GHz and 73 GHz in both LOS and NLOS channel conditions for the InH scenario and comparisons are made with the 3GPP path loss model.
\end{itemize}
\par 
\par The organization of this paper is as follows: Section II provides an overview of the measurements conducted by NYU WIRELESS. Section III describes the equivalence between the large-scale path loss model specified in 3GPP TR 38.901 for the InH scenario for both LOS and NLOS conditions and the FI and ABG path loss models for single and multi-frequencies. Section IV compares the parameters of the FI path loss model with those of the 3GPP TR 38.901 path loss model and validates the latter. Section V presents the ABG path loss model parameters for the frequency ranges of 7-24 GHz and 0.5-100 GHz. Furthermore, V.A compares the ABG path loss model parameters with the 3GPP TR 38.901 path loss model and validates the latter for the 7-24 GHz frequency range, while Section V.B performs a similar comparison for the 0.5-100 GHz frequency range. The paper concludes in Section VI by summarizing key findings, limitations and outlining directions for future research.

\section{Overview of Measurement}\label{sec:overviewMeas}
NYU WIRELESS performed real-world measurements in the InH scenario for both LOS and NLOS channel conditions at 6.75 GHz, 16.95 GHz, 28 GHz, and 73 GHz using a 1 GHz wideband sliding correlation based channel sounder \cite{shakya:2024:comprehensive-fr1,shakya:2024:propagation,maccartney:2015:indoor-office-wideband}. Specifically, the measurements at 6.75 GHz and 16.95 GHz were performed at the NYU WIRELESS Research Center, located at 370 Jay Street, Brooklyn, NY \cite{shakya:2024:comprehensive-fr1,shakya:2024:propagation}. 
The measurements at both 6.75 GHz and 16.95 GHz were conducted in identical locations. 20 distinct pairs of transmitter-receiver (TX-RX) locations were measured at each frequency (7 LOS and 13 NLOS), with separation distances of TX-RX ranging from 13 m to 97 m, and the height of TX and RX was fixed at 2.4 m and 1.5 m, respectively 
\cite{shakya:2024:comprehensive-fr1,shakya:2024:propagation}. In contrast, measurements at 28 GHz and 73 GHz were performed at the older NYU WIRELESS Research Center, located at the 2 MetroTech Center in Brooklyn, NY \cite{maccartney:2015:indoor-office-wideband}. 
The measurements at both 28 GHz and 73 GHz were conducted in identical locations. 48 distinct pairs of TX-RX locations were measured at each frequency (10 LOS and 38 NLOS), with separation distances of TX-RX ranging from 3.9 to 45.9 m and the height of TX and RX was fixed at 2.5 m and 1.5 m, respectively \cite{maccartney:2015:indoor-office-wideband}. 
The omnidirectional path loss data for 6.75 GHz and 16.95 GHz were obtained from \cite{rappaport:2024:point-data-for-site-specific, shakya:2024:comprehensive-fr1, 3gpp:2024:views-on-channel-model-sharp}, while the data for 28 GHz and 73 GHz were sourced from \cite{maccartney:2015:indoor-office-wideband}.

\section{Large-scale path loss models}\label{sec:lspl}
Path loss models are crucial for quantifying signal attenuation over distance, frequency, or both. The path loss in dB for the ABG path loss model for any scenario in both LOS and NLOS channel conditions is denoted by \eqref{abg} \cite{maccartney:2015:indoor-office-wideband}.
\begin{align}
    \label{abg}
    \mathrm{PL^{ABG}} = \mathrm{\beta} + \mathrm{10\alpha log_{10}(d_{3D})} + \mathrm{10\gamma log_{10}(f_{c})} + \mathrm{\sigma_{SF}}
\end{align}
where, where $\alpha$ and $\gamma$ are represent the path loss dependence on distance and frequency, respectively. $\beta$ is an optimized offset parameter in dB that is devoid of physical meaning, $f_{c}$ is the center frequency in GHz, $d_{3D}$ denotes the 3D TX-RX separation distance in meters and $\sigma_{SF}$ is a Gaussian random variable representing shadow fading about the distance dependent mean path loss value. Furthermore, when considering a single frequency, the ABG path loss model reverts to the FI path loss model (when setting $\gamma$ = 0 or 2 in the ABG path loss model in \eqref{abg} \cite{maccartney:2015:indoor-office-wideband}). The path loss in dB for the FI path loss model for any scenario in both LOS and NLOS
channel conditions is shown in \eqref{fi} \cite{maccartney:2015:indoor-office-wideband}. 
\begin{align}
    \label{fi}
    \mathrm{PL^{FI}} = \mathrm{\alpha} + \mathrm{10\beta log_{10}(d_{3D})} + \mathrm{\sigma_{SF}}
\end{align}
where, $\alpha$ is the floating-intercept in dB, $\beta$ denotes the path loss dependence on distance, $d_{3D}$ denotes the 3D TX-RX separation distance in meters and $\sigma_{SF}$ is a Gaussian random variable representing the shadow fading about the distance dependent mean path loss value.  A detailed explanation on how to determine the values of unknown coefficients in \eqref{abg} and \eqref{fi} based on measurement data is provided in \cite{maccartney:2015:indoor-office-wideband}.
\par Similarly, the path loss in dB for the InH scenario in both the LOS and NLOS channel conditions, as specified in 3GPP TR 38.901, are given in \eqref{3gppInHLos} and \eqref{3gppInHNLos}, respectively.
\begin{align}
    \label{3gppInHLos}
    \mathrm{PL_{{LOS}}} =  \mathrm{32.4} &+ \mathrm{17.3log_{10}(d_{3D})} + \mathrm{20log_{10}(f_{c})} \nonumber \\&+ \mathrm{\sigma_{SF}}, \hspace{0.2em}where \hspace{0.4em}\mathrm{\sigma_{SF} = 3 \hspace{0.2em}dB}
\end{align}

\begin{align}
    \label{3gppInHNLos}
    &\mathrm{PL_{NLOS}} = \mathrm{max(PL_{LOS},PL'_{NLOS})}, \nonumber\\
     \mathrm{Option1} &\mathrm{:PL'_{{NLOS}}} = 17.3 + \mathrm{38.3log_{10}(d_{3D})} + \mathrm{24.9log_{10}(f_{c})} \nonumber \\&\hspace{7.2em}+ \mathrm{\sigma_{SF}}, \hspace{0.2em}where \hspace{0.4em}\mathrm{\sigma_{SF} = 8.03 \hspace{0.2em}dB} \nonumber \\
     \mathrm{Option2}&\mathrm{:PL'_{{NLOS}}} = 32.4 + \mathrm{31.9log_{10}(d_{3D})} + \mathrm{20log_{10}(f_{c})} \nonumber \\&\hspace{7.2em}+ \mathrm{\sigma_{SF}}, \hspace{0.2em}where \hspace{0.4em}\mathrm{\sigma_{SF} = 8.29 \hspace{0.2em}dB}
\end{align}
 where $d_{3D}$ denotes the 3D TX-RX separation distance in meters, and  1 m $\leq$ $d_{3D}$ $\leq$ 150 m. $f_{c}$ denotes the center frequency in GHz, and  0.5 GHz $\leq$ $f_{c}$ $\leq$ 100 GHz. On comparing equations \eqref{3gppInHLos} and \eqref{3gppInHNLos} with \eqref{abg}, it is evident that the 3GPP TR 38.901 path loss model has a form similar to that of the ABG path loss model. Hence, the 3GPP TR 38.901 path loss model accounts for both the dependence of the path loss on distance and frequency. Similarly, on comparing equations \eqref{3gppInHLos} and \eqref{3gppInHNLos} with \eqref{fi} for a single frequency the 3GPP TR 38.901 path loss model effectively reverts to the FI path loss model. \par Due to varying measurement environments, systems and techniques, individual organizations may observe differences in measured path loss values and those predicted by 3GPP TR 38.901 path loss model for specific frequencies. This is because the 3GPP TR 38.901 path loss model spans the frequency range of 0.5-100 GHz, making it challenging to accurately tune it for specific frequencies. While combining the path loss data from different organizations in the frequency range of 7-24 GHz to analyze discrepancies compared to 3GPP TR 38.901 path loss model may be considered, significant discrepancies would still require a broader analysis across the entire 0.5-100 GHz frequency range. This is because tuning the 3GPP TR 38.901 path loss model just based on measurement data for the 7-24 GHz frequency range could introduce inaccuracies in path loss predictions for FR1 and FR2 frequencies. Hence, before making any updates to the 3GPP TR 38.901 path loss model, data from all frequency bands should be considered to ensure frequency continuity and overall model integrity for the entire frequency range of 0.5-100 GHz \cite{3gpp:2023:sid-channel-modelling}.
 \par To illustrate this concept, in the following sections we present the FI path loss model parameters at 6.75 \footnote{6.75 GHz is the center frequency of the channel sounder with a bandwidth of 1 GHz, effectively covering the frequency range of 6.25-7.25 GHz. The wideband sounder averages over this frequency range. Given the proximity of 6.75 GHz to 7 GHz, the propagation differences are negligible, and thus 6.75 GHz can be representative of 7 GHz. The center frequency of 6.75 GHz was chosen due to hardware limitations of the channel sounder at NYU WIRELESS and FCC license.} GHz and 16.95 GHz and the ABG path loss model parameters for the frequency ranges of 7-24 GHz and 0.5-100 GHz. Moreover, to validate the 3GPP TR 38.901 path loss model, we compare its parameters to those of the FI and ABG path loss models as the 3GPP TR 38.901 path loss model is equivalent to the FI and ABG path loss models for single and multi-frequencies. Additionally, the parameters for the close-in (CI) path loss model with a reference distance of 1 m for 6.75 GHz, 16.95 GHz, 28 GHz, and 73 GHz are available in \cite{shakya:2024:comprehensive-fr1,shakya:2024:propagation, maccartney:2015:indoor-office-wideband} and the FI and ABG path loss model parameters for 28 GHz and 73 GHz are provided in \cite{maccartney:2015:indoor-office-wideband}.

\begin{table*}[htbp]
  \centering
  \caption{\centering FI path loss model parameters for measured data and 3GPP TR 38.901 path loss model (\eqref{3gppInHLos},\eqref{3gppInHNLos}) for the InH scenario in LOS and NLOS environment (Env.) at 6.75 GHz and 16.95 GHz. The NLOS FI path loss model parameters for 3GPP TR 38.901 are shown for both options (Option1/Option2) in \eqref{3gppInHNLos}. |$\Delta\beta$| and |$\Delta\sigma_{SF}$| represent the absolute difference of the measured and 3GPP TR 38.901 values for the path loss dependence on distance and shadow fading.}
    \begin{tabular}{|c|c|c|c|c|c|c|c|c|c|}
    \hline
    \multirow{3}[4]{*}{Frequency} & \multirow{3}[4]{*}{Env.} & \multicolumn{3}{c|}{Measured data} & \multicolumn{3}{c|}{3GPP TR 38.901} & \multicolumn{1}{c|}{\multirow{3}[4]{*}{|$\Delta\beta$|}} & \multicolumn{1}{c|}{\multirow{3}[4]{*}{|$\Delta\sigma_{SF}$|}} \\
\cline{3-8}          &       & \multicolumn{1}{c|}{\multirow{2}[2]{*}{$\alpha$}} & \multicolumn{1}{c|}{\multirow{2}[2]{*}{$\beta$}} & \multicolumn{1}{c|}{\multirow{2}[2]{*}{$\sigma_{SF}$}} & \multicolumn{1}{c|}{\multirow{2}[2]{*}{$\alpha$}} & \multicolumn{1}{c|}{\multirow{2}[2]{*}{$\beta$}} & \multicolumn{1}{c|}{\multirow{2}[2]{*}{$\sigma_{SF}$}} &       &        \\
          &       &       &       &       &       &       &       &       &        \\
    \hline
    \multirow{2}[4]{*}{6.75 GHz} & LOS   &  43.4     &   1.7    &  3.4     &   48.98    &   1.73    &   3    &   0.03    &    0.4    \\
\cline{2-10}          & NLOS  &   35.2    &   3.6    &   9.0   &   37.94/48.98    &   3.83/3.19    &  8.03/8.29     &  0.23/0.41     &   0.97/0.71     \\
    \hline
    \multirow{2}[4]{*}{16.95 GHz} & LOS   &   50.9    &   1.7    &   2.4    &   56.98    &  1.73     &   3    &   0.03    &   0.6      \\
\cline{2-10}          & NLOS  &   61.0    &   2.8    &   8.1    &   47.90/56.98    &   3.83/3.19    &   8.03/8.29    &   1.03/0.39    &  0.07/0.19     \\
    \hline
    \end{tabular}%
  \label{tab:fiParams}%
\end{table*}%

\section{FI Path Loss Models}\label{sec:sfplm}
\begin{figure}[htbp]
    \centering
    \begin{subfigure}[b]{0.5\textwidth}
    \centering
        \includegraphics[width=9cm,height=9cm,keepaspectratio]{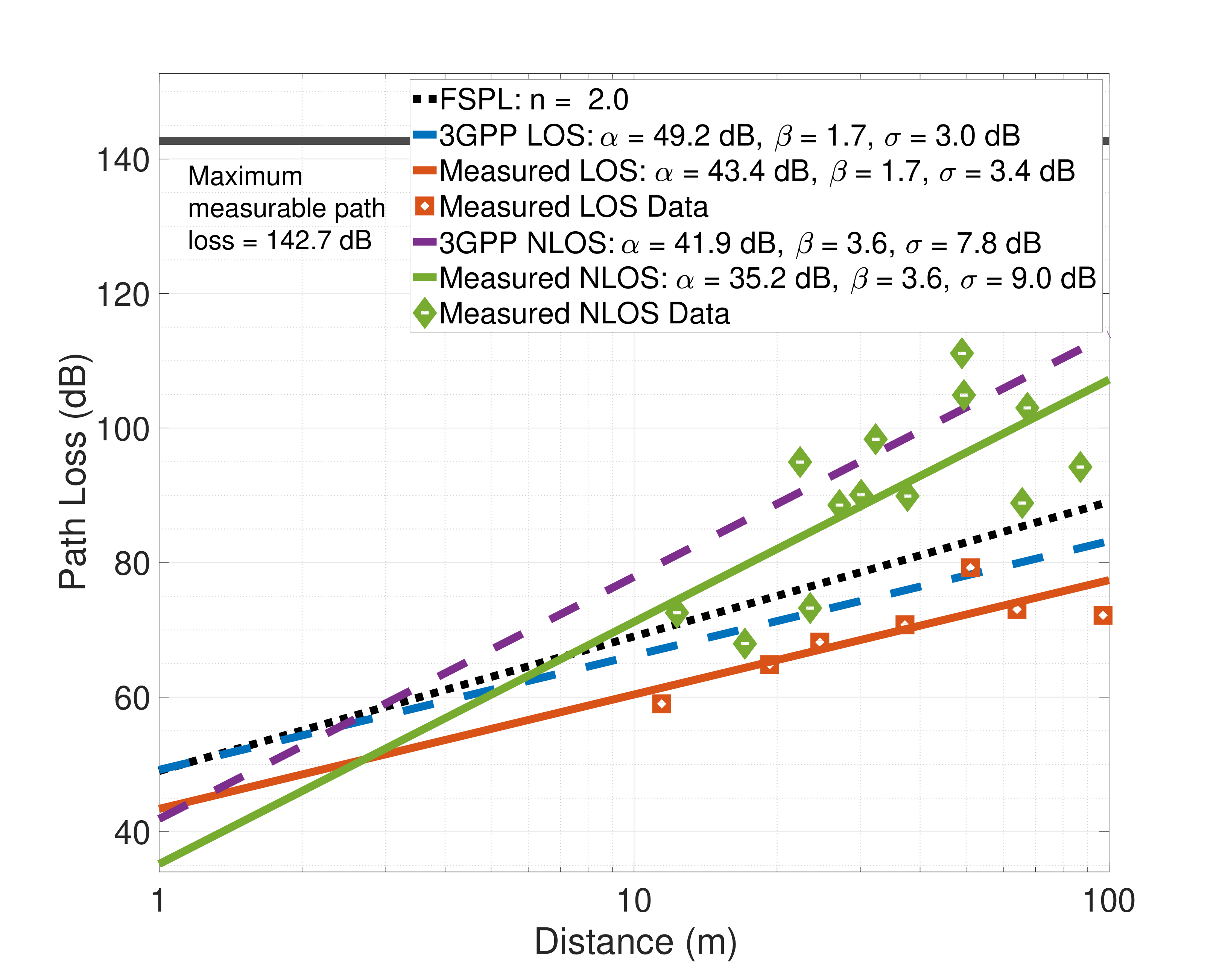}
       \centering \caption{6.75 GHz V-V InH omnidirectional path loss.}
        \label{fig:fi_6.75}
    \end{subfigure}
    \begin{subfigure}[b]{0.5\textwidth}
    \centering
        \includegraphics[width=9cm,height=9cm,keepaspectratio]{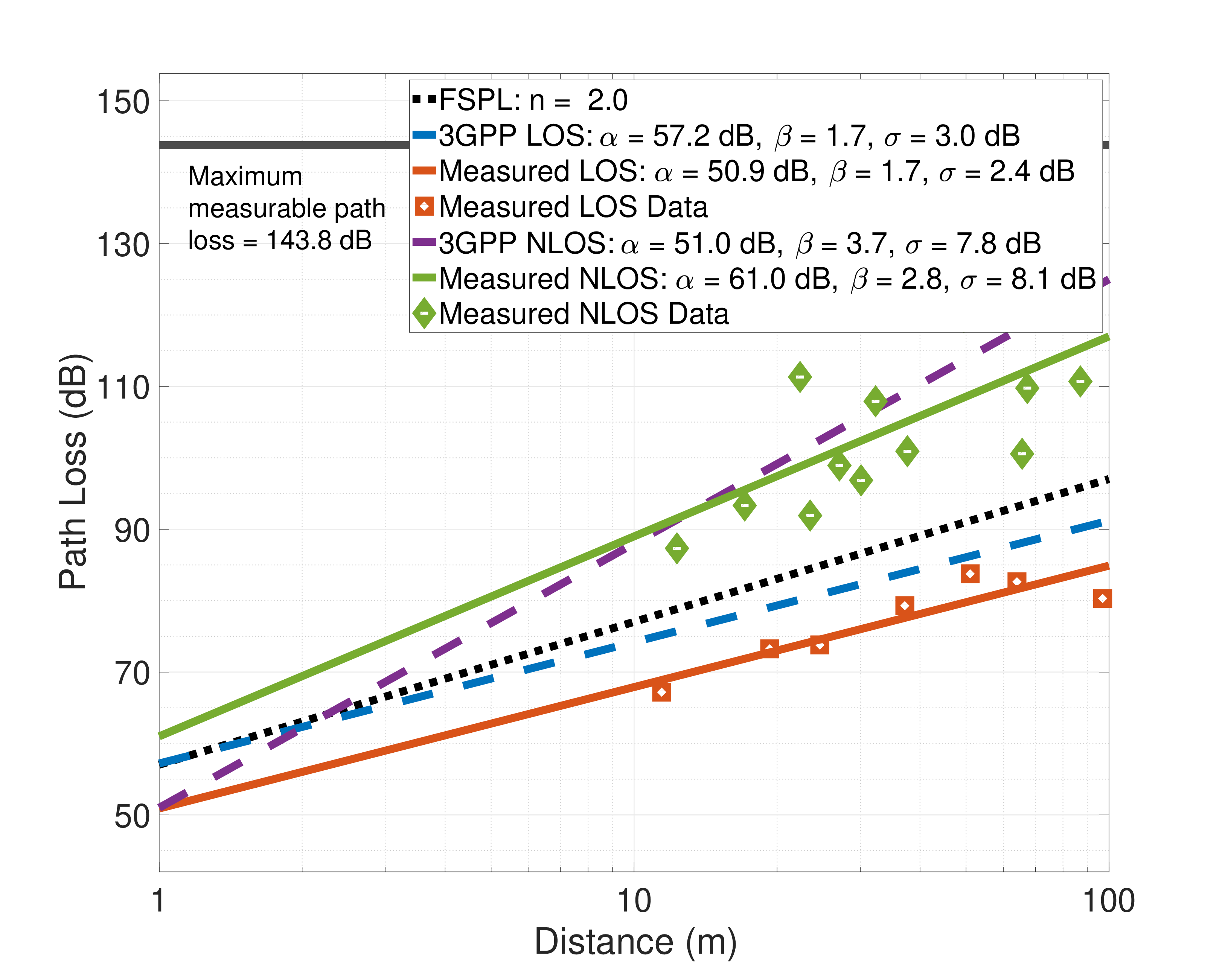}
        \caption{16.95 GHz V-V InH omnidirectional path loss.}
        \label{fig:fi_16.95}
    \end{subfigure}
    \caption{InH FI path loss scatter plots and models in both LOS and NLOS channel conditions for measured data and 3GPP path loss model over a distance range of 1 m to 100 m for V-V polarization. The NLOS FI path loss model fit for 3GPP is shown for Option1 in \eqref{3gppInHNLos}. Orange squares and green diamonds represent the measured LOS and NLOS omnidirectional path loss values.}
    \vspace{-0.2in}
    \label{fig:throughput_1}
\end{figure}
The FI path loss model provides an understanding of signal attenuation over distance at a given frequency. To validate the 3GPP TR 38.901 path loss model for the InH scenario in both LOS and NLOS channel conditions for specific frequencies in the 7-24 GHz frequency range we use the FI path loss model. Additionally, the 3GPP TR 38.901 path loss model for InH scenario in NLOS channel condition provides two options for determining the path loss, as shown in \eqref{3gppInHNLos}. Therefore, while comparing the FI path loss model parameters derived for the measured data with those from the 3GPP TR 38.901 path loss model in NLOS channel condition, we use both options.\par Fig. \ref{fig:fi_6.75} and Fig. \ref{fig:fi_16.95} present the FI path loss model fit on both the measured data and the 3GPP TR 38.901 path loss model in both LOS and NLOS channel conditions for the InH scenario at 6.75 GHz and 16.95 GHz. For readability and clarity, both the figures display only the FI path loss model fit for Option1 \eqref{3gppInHNLos} in NLOS channel condition for 3GPP TR 38.901. To perform the FI path loss model fit for the 3GPP TR 38.901 path loss model in both LOS and NLOS channel conditions, we generated samples using \eqref{3gppInHLos} and \eqref{3gppInHNLos} over the measured distances ranging from 1 m to 100 m at both frequencies. To determine the FI path loss model parameters, $\alpha$, $\beta$, and $\sigma_{SF}$ in \eqref{fi} for the measured data and 3GPP TR 38.901 path loss model, at both frequencies, we use the closed-form optimized solutions provided in Appendix A of \cite{maccartney:2015:indoor-office-wideband}.

\par Moreover, from Table \ref{tab:fiParams}, we observe that at both frequencies in LOS, the value of $\beta$ for the measured data and the 3GPP TR 38.901 path loss model is 1.7 and 1.73. respectively. In contrast, in NLOS at 6.75 GHz, the $\beta$ value for the measured data is 3.6, while for the 3GPP TR 38.901 path loss model it is 3.83 for Option1 and 3.19 for Option2. Similarly, at 16.95 GHz in NLOS, the $\beta$  value for the measured data is 2.8, while for the 3GPP TR 38.901 path loss model, it is 3.83 for Option1 and 3.19 for Option2. Thus, we observe that the measured data and the 3GPP TR 38.901 path loss model exhibit similar $\beta$ values at both frequencies in LOS (the difference between the measured value and 3GPP TR 38.901 path loss model is only 0.03 as shown in Table \ref{tab:fiParams}). Similarly, in NLOS, the $\beta$ value at 6.75 GHz for the measured data closely matches that of Option1 from the 3GPP TR 38.901 path loss model (the difference between the measured value and 3GPP TR 38.901 path loss model is 0.23 as shown in Table \ref{tab:fiParams}). At 16.95 GHz in NLOS, the $\beta$ value for the measured data is similar to Option2 (the difference between the measured value and 3GPP TR 38.901 path loss model is 0.39 as shown in Table \ref{tab:fiParams}), indicating reasonable agreement between the measured data and the 3GPP TR 38.901 path loss model at both frequencies in NLOS. Similarly, from Table \ref{tab:fiParams} we observe that the difference in shadow fading between the measured data and 3GPP TR 38.901 path loss model is less than 1 dB at both frequencies in both LOS and NLOS channel conditions. This indicates that the FI path loss model curve fitting error on measured data is minimal ($<$ 1 dB), and the overall shadow fading values obtained for the measurement data closely align with the 3GPP TR 38.901 path loss model. 

\begin{table*}[htbp]
  \centering
  \caption{\centering ABG path loss model parameters for measured data and 3GPP TR 38.901 path loss model (\eqref{3gppInHLos},\eqref{3gppInHNLos}) for the InH scenario in LOS and NLOS environment (Env.) for 7-24 GHz and 0.5-100 GHz frequency (Freq.) range. The NLOS ABG path loss model parameters for 3GPP TR 38.901 is shown for both options (Option1/Option2) in \eqref{3gppInHNLos}. |$\Delta\alpha$|, |$\Delta\gamma$| and |$\Delta\sigma_{SF}$| represent the absolute difference of the measured and 3GPP TR 38.901 values for the path loss dependence on distance, frequency and shadow fading.}
    \begin{tabular}{|p{2.8em}|c|c|c|c|c|c|c|c|c|p{3.2em}|p{3.2em}|p{3.2em}|}
    \hline
    \multirow{3}[3]{*}{\hspace{0.5em}Freq.} & \multirow{3}[3]{*}{Env.} & \multicolumn{4}{c|}{Measured data} & \multicolumn{4}{c|}{3GPP TR 38.901} & \multirow{3}[3]{*}{\hspace{0.8em}|$\Delta\alpha$|} & \multirow{3}[3]{*}{\hspace{0.8em}|$\Delta\gamma$|} & \multirow{3}[3]{*}{|$\Delta\sigma_{SF}$|}\\
\cline{3-10} &  & \multirow{2}[2]{*}{$\alpha$} & \multirow{2}[2]{*}{$\beta$} & \multirow{2}[2]{*}{$\gamma$} & \multirow{2}[2]{*}{$\sigma_{SF}$} & \multirow{2}[2]{*}{$\alpha$} & \multirow{2}[2]{*}{$\beta$} & \multirow{2}[2]{*}{$\gamma$} & \multirow{2}[2]{*}{$\sigma_{SF}$} &  &  &  \\
    &       &       &       &       &       &       &       &       &       &       &   &\\
    \hline
    \multirow{2}[4]{*}{\hspace{0.5em}7-24} & LOS   &    1.7   &   28.2    &    1.9   &   2.9    & 1.73       &    32.4   &   2    &  3 & \hspace{0.8em}0.03 & \hspace{0.8em}0.1 & \hspace{0.8em}0.1\\
\cline{2-13}          & NLOS  &    3.2   &   12.9    &   3.4    &   8.6    &  3.83/3.19     &   17.3/32.4    &    2.49/2   &  8.03/8.29 & 0.63/0.01 & 0.91/1.4 & 0.57/0.31\\
    \hline
    \multirow{2}[4]{*}{0.5-100} & LOS   &   1.4    &   29.5    &    2.1   &  2.7     & 1.73       &    32.4   &   2    &  3 & \hspace{0.8em}0.33 & \hspace{0.8em}0.1 & \hspace{0.8em}0.3\\
\cline{2-13}          & NLOS  &    3.4   &    12.9   &   2.9    &    10.1   &  3.83/3.19     &   17.3/32.4    &    2.49/2   &  8.03/8.29 & 0.43/0.21 &0.41/0.9 & 2.07/1.81\\
    \hline
    \end{tabular}%
  \label{tab:abgParams}%
\end{table*}%

\section{ABG Path Loss Models}\label{sec:mfplm}
The ABG path loss model provides valuable insights into both the path loss dependence on distance and frequency. To obtain the parameters of the ABG path loss model for the 7-24 GHz frequency range, we combine the measurement data obtained for the InH scenario in the LOS channel condition at 6.75 GHz and 16.95 GHz and the NLOS channel condition at 6.75 GHz and 16.95 GHz. Similarly, to derive the ABG path loss model parameters for the frequency range of 0.5–100 GHz, we combine the measurement data obtained for the InH scenario in LOS channel condition at 6.75 GHz, 16.95 GHz, 28 GHz and 73 GHz and NLOS channel condition at 6.75 GHz, 16.95 GHz, 28 GHz and 73 GHz. To validate the 3GPP TR 38.901 path loss model for the InH scenario in both LOS and NLOS channel conditions for 7-24 GHz and 0.5-100 GHz frequency ranges, we use the derived ABG path loss model parameters from the measured data. To determine the ABG path loss model parameters, $\alpha$, $\beta$, $\gamma$ and $\sigma_{SF}$ in \eqref{abg}, we use the closed-form optimized solutions provided in Appendix A of \cite{maccartney:2015:indoor-office-wideband}. Furthermore, in the NLOS channel condition, the 3GPP TR 38.901 path loss model provides two options to determine the path loss, as shown in \eqref{3gppInHNLos}. Therefore, while comparing the ABG path loss model parameters derived from the measured data with those from the 3GPP TR 38.901 path loss model, we use both options.
\subsection{ABG Path Loss Model Parameters for 7-24 GHz}
\begin{figure}[h!]
    \centering
    \begin{subfigure}[b]{0.5\textwidth}
    \centering
        \includegraphics[width=8.9cm,height=8.9cm,keepaspectratio]{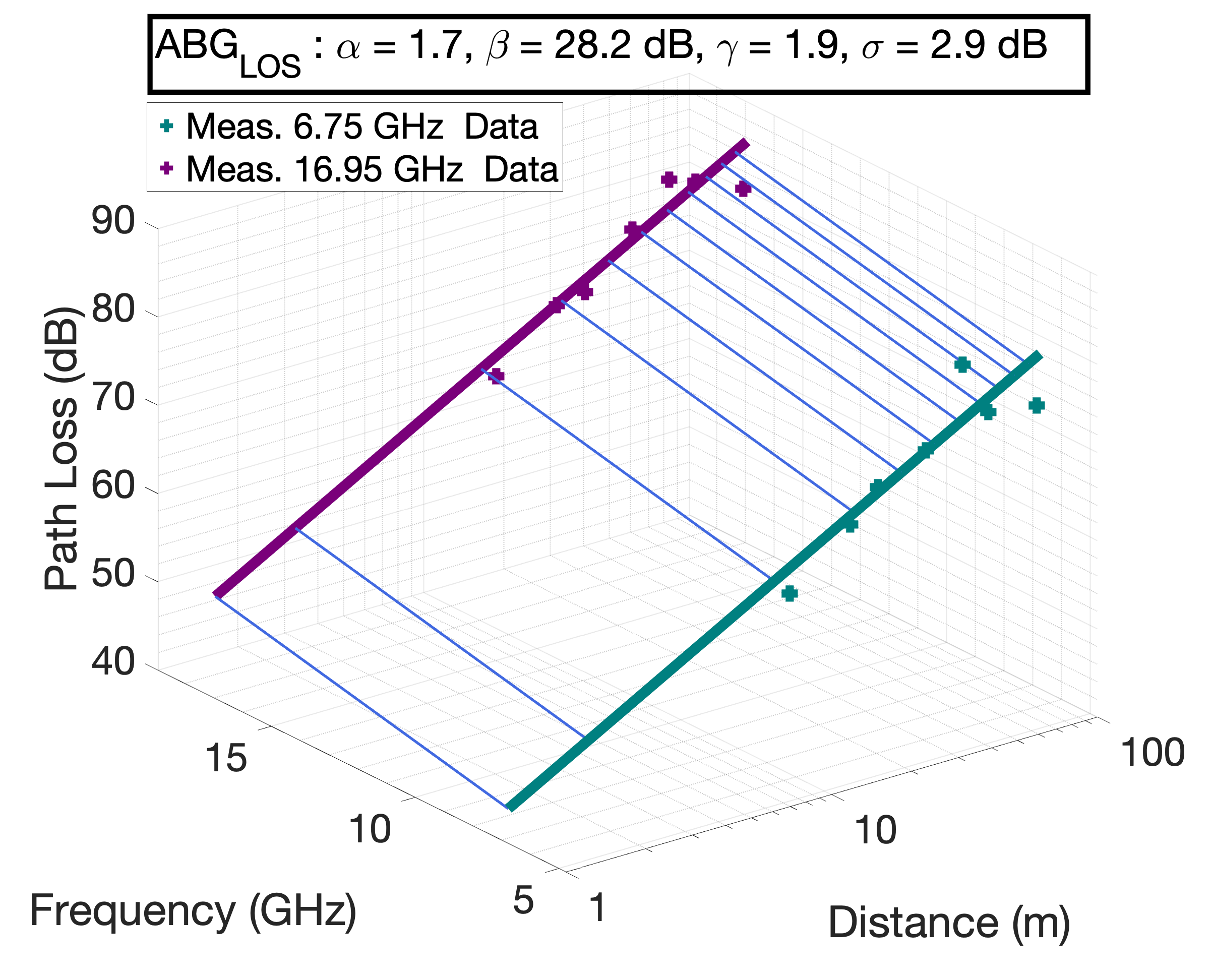}
       \centering \caption{7-24 GHz V-V InH omnidirectional LOS path loss.}
        \label{fig:fr3los}
    \end{subfigure}
    \begin{subfigure}[b]{0.5\textwidth}
    \centering
        \includegraphics[width=8.9cm,height=8.9cm,keepaspectratio]{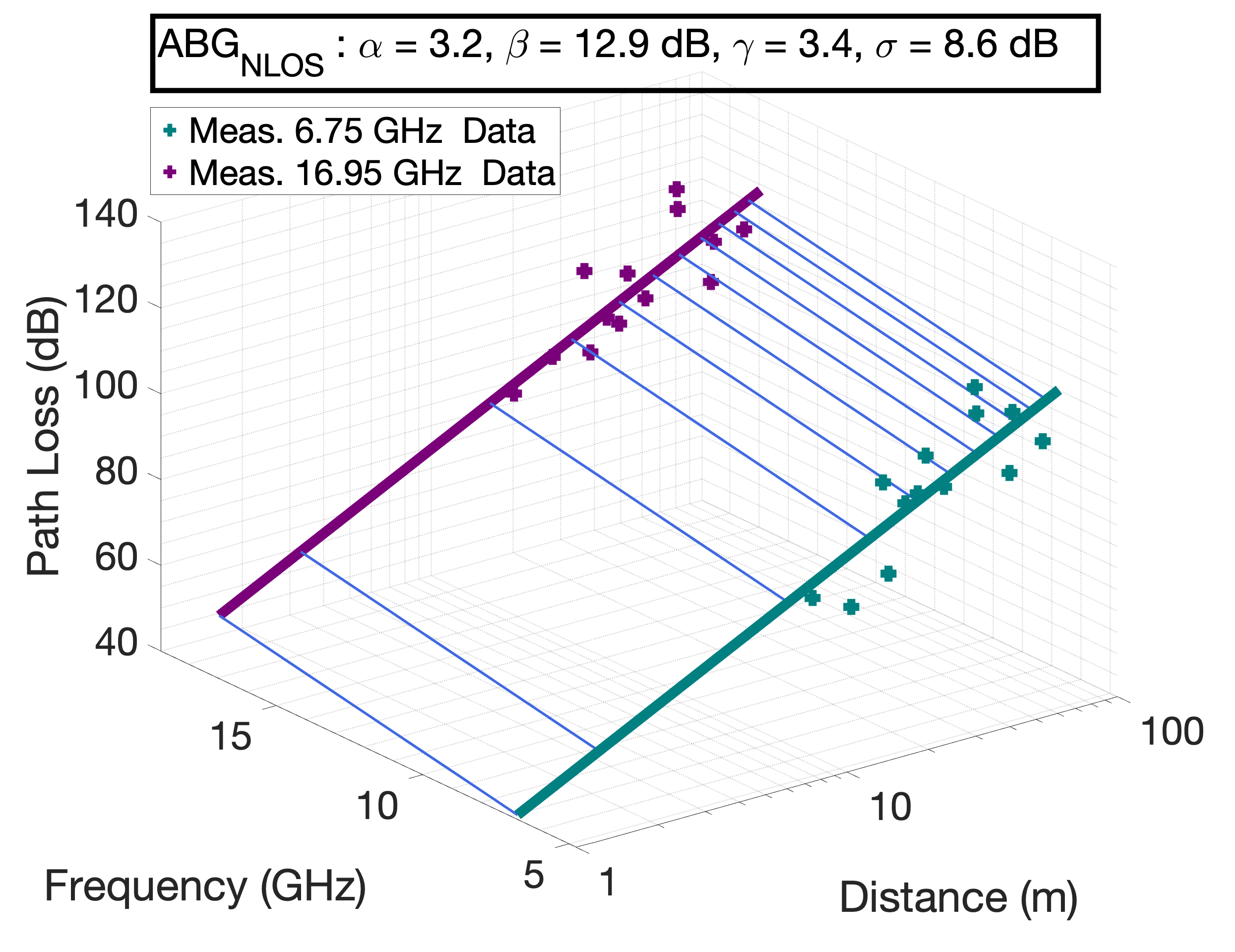}
        \caption{7-24 GHz V-V InH omnidirectional NLOS path loss.}
        \label{fig:fr3nlos}
    \end{subfigure}
    \caption{InH ABG path loss scatter plots and models for measured data  over a distance range of 1 m to 100 m for V-V polarization.}
    
    \label{fig:xx}
\end{figure}
Fig. \ref{fig:fr3los} and Fig. \ref{fig:fr3nlos} present the ABG path loss model fit on the measured data in both LOS and NLOS channel conditions for the frequency range of 7-24 GHz. From Table \ref{tab:abgParams}, in LOS, we can see that the value of $\alpha$ for the measured data and 3GPP TR 38.901 path loss model is 1.7 and 1.73, respectively. Furthermore, the value of $\gamma$ for the measured data is 1.9 and the 3GPP TR 38.901 path loss model is 2. In contrast, in NLOS, the value of $\alpha$ for the measured data is 3.2, while for the 3GPP TR 38.901 path loss model it is 3.83 for Option1 and 3.19 for Option2. Furthermore, the value of $\gamma$ for the measured data is 3.4, while for the 3GPP TR 38.901 path loss model it is 2.49 for Option1 and 2 for Option2. Thus, we observe that the measured data and the 3GPP TR 38.901 path loss model exhibit similar $\alpha$ values in LOS (the difference between the measured value and 3GPP TR 38.901 path loss model is 0.03 as shown in Table \ref{tab:abgParams}). On the other hand, the value of $\alpha$ in NLOS is similar to that of the 3GPP TR 38.901 path loss model Option2 (the difference between the measured value and 3GPP TR 38.901 path loss model is 0.01 as shown in Table \ref{tab:abgParams}). Similarly, the value of $\gamma$ in LOS is similar to the 3GPP TR 38.901 path loss model and in NLOS the value of $\gamma$ has some discrepancy compared to the 3GPP TR 38.901 path loss model (the difference between the measured value and 3GPP TR 38.901 path loss model Option1 and Option2 are 0.91 and 1.4, respectively as shown in Table \ref{tab:abgParams}). The observed discrepancy for $\gamma$ may stem from the limited number of frequency points (only two). More measurements at different frequencies in 7-24 GHz band are needed to accurately capture the frequency dependence of path loss in NLOS conditions. Notably, a lower $\gamma$ in 3GPP TR 38.901 compared to measurements suggests that the 3GPP path loss model underestimates the path loss at higher frequencies for a given distance within the 7–24 GHz band. However, since the 3GPP model spans a wide frequency range (0.5–100 GHz), refining it based solely on 7–24 GHz data could lead to overfitting. Thus, before any refinements are made, data from the entire frequency range should be considered. Moreover, we can also see from Table \ref{tab:abgParams} that the difference in shadow fading between the measured data and 3GPP TR 38.901 path loss model is less than 0.6 dB for the frequency range of 7-24 GHz in both LOS and NLOS channel conditions. This indicates that the ABG path loss model curve fitting error on measured data is negligible (< 0.6 dB), and the overall shadow fading values obtained for measurement data closely align with the 3GPP TR 38.901 path loss model.

\subsection{ABG Path Loss Model Parameters for 0.5-100 GHz}
\begin{figure}[h!]
    \centering
    \begin{subfigure}[b]{0.5\textwidth}
    \centering
        \includegraphics[width=8.9cm,height=8.9cm,keepaspectratio]{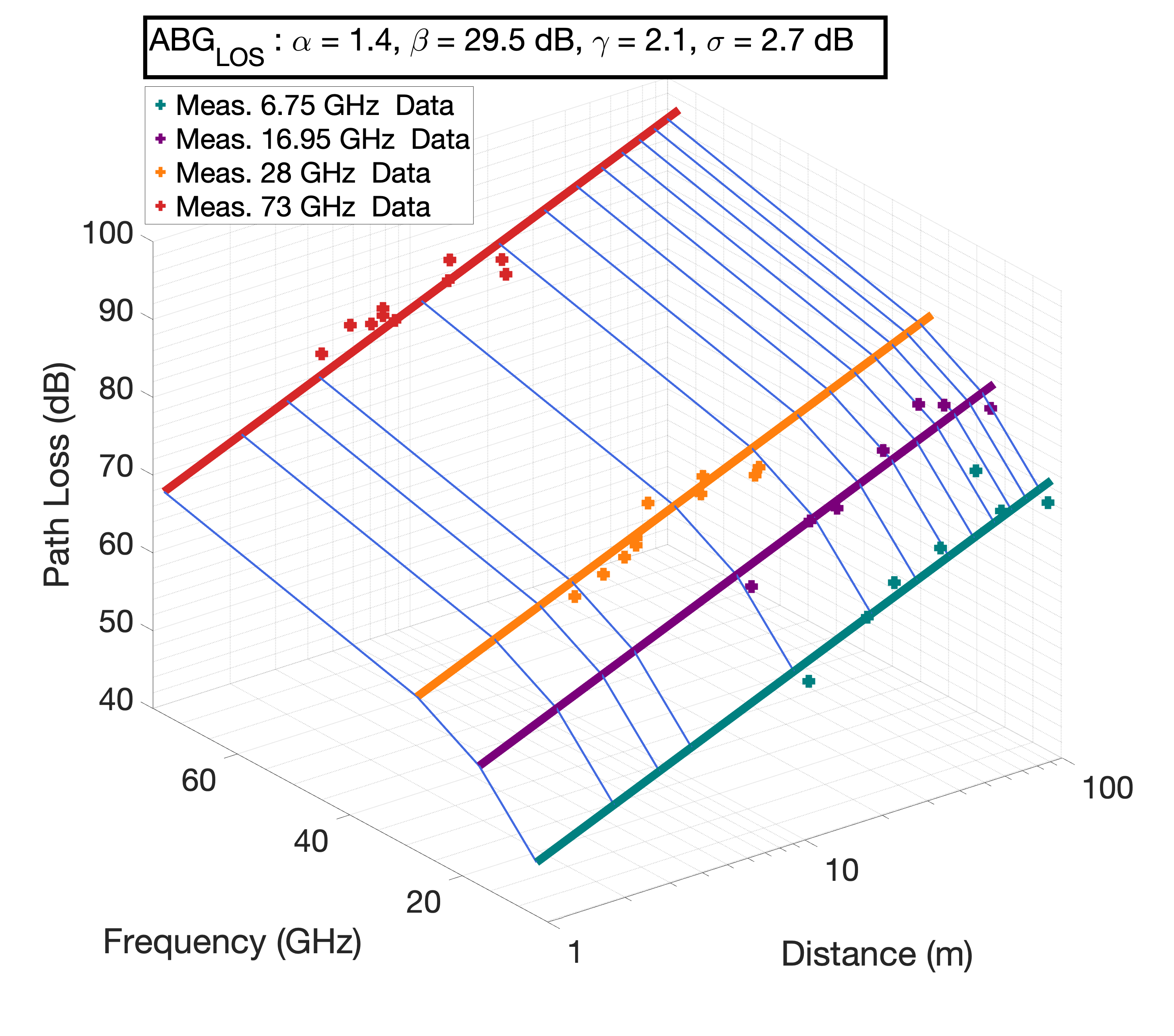}
       \centering \caption{0.5-100 GHz V-V InH omnidirectional LOS path loss.}
        \label{fig:overalllos}
    \end{subfigure}
    \begin{subfigure}[b]{0.5\textwidth}
    \centering
        \includegraphics[width=8.9cm,height=8.9cm,keepaspectratio]{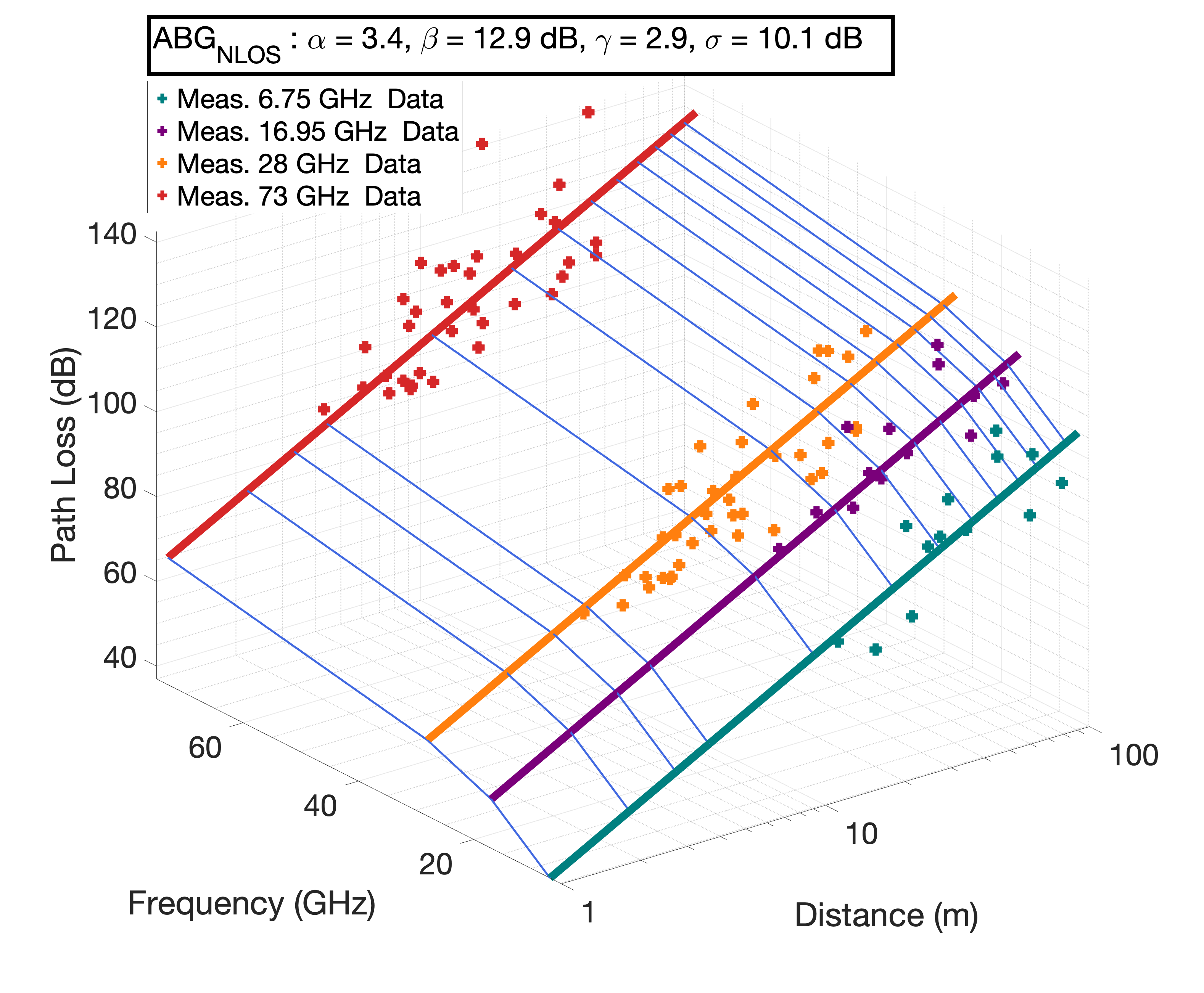}
        \caption{0.5-100 GHz V-V InH omnidirectional NLOS path loss.}
        \label{fig:overallnlos}
    \end{subfigure}
    \caption{InH ABG path loss scatter plots and models for measured data  over a distance range of 1 m to 100 m for V-V polarization.}
    \vspace{-0.2in}
    \label{fig:xx}
\end{figure}

Fig. \ref{fig:overalllos} and Fig. \ref{fig:overallnlos} present the ABG path loss model fit on the measured data in both LOS and NLOS channel conditions for the frequency range of 0.5-100 GHz. 
From Table \ref{tab:abgParams}, in LOS, we observe that the value of $\alpha$ for the measured data and 3GPP TR 38.901 path loss model is 1.4 and 1.73, respectively. Furthermore, the value of $\gamma$ for the measured data is 2.1 and the 3GPP TR 38.901 path loss model is 2. In contrast, in NLOS, the value of $\alpha$ for the measured data is 3.4, while for the 3GPP TR 38.901 path loss model it is 3.83 for Option1 and 3.19 for Option2. Furthermore, the value of $\gamma$ for the measured data is 2.9, while for the 3GPP TR 38.901 path loss model it is 2.49 for Option1 and 2 for Option2. 
Thus, we observe that the measured data and the 3GPP TR 38.901 path loss model exhibit similar $\alpha$ value in LOS (the difference between the measured value and 3GPP TR 38.901 path loss model is 0.33 as shown in Table \ref{tab:abgParams}). On the other hand, the value of $\alpha$ in NLOS is similar to that of the 3GPP TR 38.901 path loss model Option2 (the difference between the measured value and 3GPP TR 38.901 path loss model is 0.21 as shown in Table \ref{tab:abgParams}). Similarly, the value of $\gamma$ in LOS is similar to the 3GPP TR 38.901 path loss model and in NLOS the value of $\gamma$ has minor discrepancy compared to 3GPP TR 38.901 path loss model Option1 (the difference between the measured value and 3GPP TR 38.901 path loss model is 0.41 as shown in Table \ref{tab:abgParams}). Compared to Section V.A, the discrepancy in $\gamma$ reduces between measured data and 3GPP TR 38.901 path loss model as the number of frequency points increases. Thus, incorporating more frequency points over the entire range of 0.5-100 GHz provides a better estimate of $\gamma$. Hence, the current 3GPP TR 38.901 path loss model is valid for the entire frequency range of 0.5-100 GHz and might not require any updates specifically for the frequency range of 7-24 GHz.


\section{Conclusion}\label{sec:conclusion}
This study provides a thorough validation of the 3GPP TR 38.901 path loss model for the InH scenario in both LOS and NLOS channel conditions. The close alignment of the FI and ABG path loss model parameters for the measured data with the 3GPP TR 38.901 path loss model demonstrates that the existing 3GPP TR 38.901 path loss model, for the InH scenario in both LOS and NLOS channel conditions, is valid for the 7-24 GHz frequency range. Using only two frequency points within the 7–24 GHz range and four across the 0.5–100 GHz spectrum makes the ABG path loss model parameters derived in this study sensitive to the specific measured data and potentially more environment-specific. This limits the robustness and generalizability of the model. Incorporating additional frequency points across diverse indoor environments would improve the reliability and broader applicability of the ABG model parameters, leading to a more accurate and comprehensive validation of the 3GPP TR 38.901 InH path loss model. Hence, in the future, researchers should incorporate measurements from different frequencies within the 7-24 GHz frequency range, from diverse indoor environments with varying structural characteristics to assess the generalizability and validity of the 3GPP TR 38.901 path loss model.


\bibliographystyle{IEEEtran}

\vspace{12pt}
\color{red}
\end{document}